**Fluctuations of Energy-Relaxation Times in Superconducting Qubits**

P. V. Klimov[1], J. Kelly[1], Z. Chen[1], M. Neeley[1], A. Megrant[1], B. Burkett[1], R. Barends[1], K. Arya[1], B. Chiaro[3], Yu Chen[1], A. Dunsworth[3], A. Fowler[1], B. Foxen[3], C. Gidney[1], M. Giustina[1], R. Graff[1], T. Huang[1], E. Jeffrey[1], Erik Lucero[1], J.Y. Mutus[1], O. Naaman[1], C. Neill[1], C. Quintana[1], P. Roushan[1], Daniel Sank[1], A. Vainsencher[1], J. Wenner[3], T.C. White[1], S. Boixo[2], R. Babbush[2], V. N. Smelyanskiy[2], H. Neven[2], John M. Martinis[1]

1. Google, Santa Barbara, California 93117
2. Google, Los Angeles, California 90291
3. University of California, Santa Barbara, California 93117

Abstract
Superconducting qubits are an attractive platform for quantum computing since they have demonstrated high-fidelity quantum gates and extensibility to modest system sizes. Nonetheless, an outstanding challenge is stabilizing their energy-relaxation times, which can fluctuate unpredictably in frequency and time. Here, we use qubits as spectral and temporal probes of individual two-level-system defects to provide direct evidence that they are responsible for the largest fluctuations. This research lays the foundation for stabilizing qubit performance through calibration, design and fabrication.

Text
Superconducting circuits are attractive candidates for implementing qubits as high-fidelity quantum gates have already demonstrated in modest system sizes [1-10]. A primary challenge in scaling such circuits into a quantum computer that can solve practical problems is not only a matter of improving their performance but also stabilizing it. In particular, it has been observed in numerous architectures that qubit energy-relaxation times ($T_1$) can fluctuate unpredictably by up to an order of magnitude in time [11-16] and in frequency [8,14,16-21]. Since $T_1$ directly limits gate fidelity, these fluctuations present an obstacle for future scalability.

In past reports, $T_1$ fluctuations were attributed to quasiparticles [8] or two-level-system (TLS) defects [11-21]. With few exceptions, these conclusions were drawn by analyzing spectrally *or* temporally resolved qubit $T_1$ data, which offer limited insight into the mechanisms driving relaxation. Here we simultaneously spectrally *and* temporally resolve qubit $T_1$ to provide direct evidence that the most significant fluctuations can be explained by TLS defects and time-dependent variations in their transition frequencies - a phenomenon known as spectral diffusion. We tentatively explain the spectral diffusion dynamics via the interacting defect model, which is consistent with our observations [14,16,19,22-25]. Interestingly, the $T_1$ distributions that we extract from spectral and temporal slices of our data are consistent with those observed in other qubit and resonator architectures [7,8,11,12,14,15,26], suggesting that similar defect physics may be at play.

Two-level-system defects have been investigated for decades and were originally used to explain the low-temperature properties of amorphous solids [27]. More recently, they have been identified as a primary source of dielectric loss in superconducting circuits [28]. The microscopic nature of TLS defects is not well understood [29], but they are believed to reside in the amorphous dielectrics present at the material interfaces of superconducting circuits and within Josephson-junctions. Defects can resonantly interact with qubits and serve as a strong energy-relaxation channel [17] (see Fig. 1).

In this report, we spectrally and temporally resolve $T_1$ of frequency-tunable Xmon transmon qubits [1-3,17,30]. The spectral data is used to identify defects and the temporal data is used to infer their dynamics. The experimental pulse sequence that we use to measure $T_1$ at a single frequency is as follows: We initialize the qubit into its |0⟩ state, excite it into |1⟩, tune it to the frequency of interest, wait a variable delay time, and then measure its state. To resolve a single T1 curve, we repeat this sequence 2000 times at each of 40 log-spaced delays from 0.01 to 100 µs. Our active initialization protocol takes 7 µs and has fidelity >0.99. Our readout protocol takes 1 µs and has fidelity >0.95. With these protocols, we can quickly resolve a $T_1$ curve at a single frequency in ~2 seconds, and a spectroscopic $T_1$ trace across 400 MHz with a 1 MHz resolution in ~15 minutes. We have verified that our qubit-frequency calibration is stable to within ∓ 1 MHz across all of our measurements [30].

A spectrally- and temporally-resolved $T_1$ dataset for a single qubit is shown in Fig 2a [30]. To better illustrate the dramatic fluctuations in $T_1$, we show linecuts at constant frequency (Fig. 2b) and constant time (Fig. 2c). We see that $T_1$ can vary by up to an order of magnitude, and fluctuations between extrema can happen abruptly on 15-minute timescales, and across 5-MHz frequencies. The $T_1$ distributions of these line cuts are presented in Figs. 2d and e. In time, the distribution can have a single- or multi-modal shape, with the latter being characteristic of telegraphic noise. In frequency, the distributions are weighted heavily near their maxima, but have long tails towards low $T_1$ due to deep but sparse relaxation resonances (Fig. 2d).

Most regions of strong $1/T_1$ relaxation are characteristic of resonant relaxation into a coupled system, such as a TLS defect [17] or an electromagnetic cavity [30-32]. Importantly, resonant relaxation is not expected for alternative mechanisms such as quasiparticles, capacitor loss, inductor loss, or radiation into a continuum [31,32]. We fit each relaxation resonance to a Lorentzian parametrized by the coupled system's relaxation and transverse coupling rates. We ascribe most resonances to defects since their respective coupling rates range from 50 - 500 kHz, which are consistent with 1 $e$Å dipole moments coupling to electric fields in the qubit capacitor or near it's Josephson junction [17,30]. Furthermore, their measured decoherence rates range from ~0.5 - 20 MHz, which are consistent with defects previously observed similar architectures [17,18, 25]. We ascribe several weak periodic resonances to modes in our qubit control lines and a sharp resonance near 5.6 GHz to bleedthrough of our microwave carrier [30]. We do not analyze the small background fluctuations but believe they can be explained by weakly coupled defects [17], quasiparticles [8], and measurement uncertainty.

To investigate spectral diffusion, we extract the center frequency of each defect's Lorentzian as a function of time, and ascribe it to that defect's transition frequency. We consolidate the transition frequencies of 13 defects across several nominally identical qubits on the same chip (Fig. 3a) and find that their standard deviation evolves in time roughly diffusively as $\sigma(t) = 2Dt^{1/2}$, with the diffusivity $D = 2.5 \mp 0.1$ MHz (hour)$^{-\frac{1}{2}}$. Nonetheless, a diffusion model by itself oversimplifies the dynamics. Interestingly, defects exhibit a combination of two distinct spectral diffusion regimes - telegraphic and diffusive. Defects in the telegraphic regime experience discrete jumps in frequency, while those in the diffusive regime experience continuous drifts (Fig. 3b). Below we investigate these dynamics.

We do not expect the TLS defects that we observe to exhibit any thermal dynamics in isolation. With transition frequencies of $E_{TLS}/h \sim 5.5$ GHz and a nominal temperature of $T = 15$ mK, their Boltzmann factors are a negligible $\exp(-E_{TLS}/k_B T) \sim 10^{-8}$. Furthermore, strain fluctuations and related defect dynamics [19, 20] should be negligible since our sample temperature is stable to within 2 mK, since the thermal expansion coefficients of all relevant materials are small at cryogenic temperatures, and since strain-defect coupling is generally weak [24]. To explain the spectral diffusion that we observe, we invoke the interacting defect model [14,16,19,22-25], in which TLS defects with $E_{TLS} \gg k_B T$ - such as those observed in our $T_1$ data - interact with thermally fluctuating defects (TF) with $E_{TF} \lesssim k_B T$. Below we introduce this model and describe how telegraphic and diffusive dynamics can emerge from it.

Each TLS and TF defect can be modeled with a tunneling Hamiltonian of the form $\hat{H} = \varepsilon \hat{\tau}_z + \Delta \hat{\tau}_x = E \hat{\sigma}_z$ [27]. Here $\varepsilon$ is the energy asymmetry of the defect's potential energy wells, $\Delta$ is the tunneling energy between them, $E = \sqrt{\varepsilon^2 + \Delta^2}$ is the transition energy between energy eigenstates, and $\hat{\tau}_i, \hat{\sigma}_i$ are the Pauli matrices in the un-diagonalized and diagonalized bases, respectively. TLS and TF defects couple to each other through the interaction Hamiltonian $\hat{H}_{int} = \frac{1}{2}\Sigma_{i,j}^{x,y,z} g_{ij} \hat{\tau}_{i,TLS} \hat{\tau}_{j,TF}$, where the coupling tensor $g_{ij}$ contains dipolar and elastic contributions that depend on the defects' structures, separation, and host material. By virtue of their drastically different energies, TLS and TF defects couple in the off-resonant limit, in which transversal coupling is negligible. The only substantial coupling term is thus $\hat{H}_{zz} = \frac{1}{2}g_{zz}\hat{\tau}_{z,TLS}\hat{\tau}_{z,TF} = \frac{1}{2}g_{\parallel}\hat{\sigma}_{z,TLS}\hat{\sigma}_{z,TF}$, where $g_{\parallel} = g_{zz}(\frac{\varepsilon_{TLS}}{E_{TLS}}\frac{\varepsilon_{TF}}{E_{TF}})$ [19,25].

To understand how spectral diffusion can emerge from this model, we inspect the energy-level structure of a coupled TLS-TF system (Fig. 3c). In this system, the single-excitation TF and TLS transition frequencies are $E_{TLS}/h = (E_{0,TLS} + g_{\parallel}\langle\hat{\sigma}_{z,TF}\rangle)/h$, and $E_{TF}/h = (E_{0,TF} + g_{\parallel}\langle\hat{\sigma}_{z,TLS}\rangle)/h$, respectively, where $E_0/h$ denotes the uncoupled frequency. From these expressions, we see that the TLS transition frequency depends on the state of its coupled TF, and vice-versa. Therefore, as the TF thermally transitions between its energy eigenstates, the TLS transition frequency jumps by $2g_{\parallel}/h$, which is observed as telegraphic spectral diffusion. The rate of telegraphic jumping is determined by the TF's phononic excitation and relaxation rates, which are $\Gamma_{e\to g} = \alpha \Delta_{TF}^2 E_{TF} \coth(E_{TF}/2k_B T)$ and $\Gamma_{g\to e} = \exp(-E_{TF}/k_B T)\Gamma_{e\to g}$, respectively. Here $\alpha$ is a constant related to the phonon-TF coupling rate, material density, and the speed of sound [23-25, 27]. Diffusive spectral diffusion is expected to emerge in the bath limit of this model, in which a single TLS is coupled to many TFs with distinct coupling and telegraphic jumping rates [16].

We now analyze our observations in the context of the interacting defect model and confirm that they are consistent [30]. The largest telegraphic jump that we observe is ~ 60 MHz, which corresponds to a TLS-TF coupling rate $g_{\parallel}/h$ = 30 MHz. This coupling magnitude makes physical sense since it corresponds to the interaction of two collinear 1 $e$Å dipoles separated by ~35 nm in a material with a relative permittivity of 10. The average telegraphic jump rates $\Gamma = (\Gamma_{g \to e} + \Gamma_{e \to g}) / 2$ that we observe range from ~ 50 uHz to 5 mHz, and they are roughly distributed as $\sim \Gamma^{-1}$. This distribution is expected for tunneling defects [24], but the range is somewhat surprising, since typical TLS relaxation rates are ~1 MHz [25]. This disparity may be explained by the quadratic scaling of $\Gamma_{e \to g}$ on the tunneling energy $\Delta$, which can vary from uHz to GHz for defects in similar materials [16, 19, 25]. The scaling of $\Gamma_{e \to g}$ on the transition energy $E$ cannot close this disparity. For several defects where many telegraphic jumps are observed, we estimate the TF energy $E_{TF}/k_B T = \ln(\Gamma_{e \to g}/\Gamma_{g \to e})$, which ranges from 0.18 to 0.99. This is consistent with our primary hypothesis that spectral diffusion is driven by thermal fluctuators.

To estimate the density of TF defects and to understand the relationship between the telegraphic and diffusive regimes, we run a Markov-Chain Monte Carlo simulation of interacting defect dynamics in a thin film representative of the interfacial dielectrics in our qubit circuit. Since the diffusivity of TLS transitions is expected to depend strongly on TF density, we use it to connect simulation to experiment. We find that our experimentally measured diffusivity $D$ = 2.5 $\mp$ 0.1 MHz (hour)$^{-\frac{1}{2}}$ is consistently reproduced by our simulation at TF densities above $10^4$ GHz$^{-1}$ µm$^{-3}$ (~25 × $10^{20}$ eV$^{-1}$ cm$^{-3}$). Furthermore, at a fixed density of ~$10^4$ GHz$^{-1}$ µm$^{-3}$, the simulation qualitatively reproduces virtually all of our diffusive and telegraphic data (see Fig. S3 in ref. 30). This density is ~10x higher than the densities typically quoted for bulk dielectrics, but this is not unexpected for interfacial thin films [22]. These simulations demonstrate that the observed spectral diffusion dynamics can emerge from the interacting defect model at a single TF density that is physically plausible.

We spectrally and temporally resolved qubit $T_1$. In these data we identified single TLS defects and tracked their spectral diffusion dynamics, which we attribute to the interacting defect model. We find that defects and their spectral diffusion are directly responsible for the most significant time- and frequency-domain qubit $T_1$ fluctuations. Interestingly, the $T_1$ distributions that we extract from time- and frequency-domain cuts of our data are qualitatively similar to those seen in in planar resonators [11,12,26], fixed-frequency 3D transmon qubits [14], fixed-frequency planar transmon qubits [7,15], and flux qubits [8]. This correspondence suggests that TLS defects may be the source of $T_1$ fluctuations in many superconducting quantum computing architectures.

Our results suggest that understanding defect properties at scale is important for stabilizing and improving the energy-relaxation times of superconducting circuits. In the short term, defect data should guide qubit calibration protocols. For example, defects' diffusivity and coherence properties should inform the algorithms that are used to select tunable qubits' frequencies and how often those algorithms are run. In the long term, defect data should guide qubit design and fabrication parameters. For example, the relationship between spectral diffusion and defect density suggests that correlation studies can be used to identify defective circuit components and materials. Ultimately, improved qubit calibration, circuit design, and fabrication will likely be necessary to manufacture and operate a quantum computer that can solve practical problems.

Acknowledgements

This work was supported by Google. C.Q. and Z.C. acknowledge support from the National Science Foundation Graduate Research Fellowship under Grant No. DGE-1144085. Devices were made at the UC Santa Barbara Nanofabrication Facility, a part of the NSF funded National Nanotechnology Infrastructure Network.

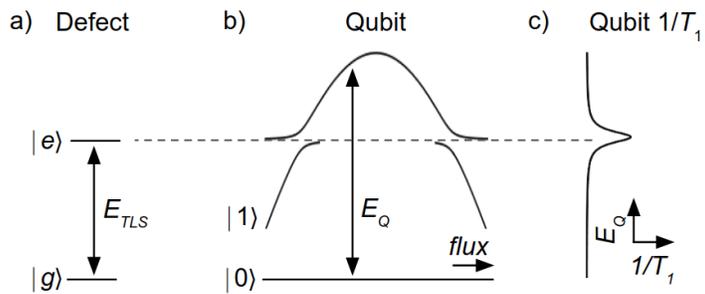

**Figure 1. Defect-Qubit Coupling.** (a) The energy-level diagram of a two-level-system defect. $E_{TLS}$ is the defect transition energy, which is defined by its physical structure. (b) The energy-level diagram of the lowest two states of a frequency-tunable Xmon transmon qubit. $E_Q$ is the qubit transition energy, which can be tuned by applying a magnetic flux to the Xmon SQUID. (c) Defects can resonantly couple to qubits and serve as a strong energy-relaxation channel with a Lorentzian spectroscopic signature. This signature is used to identify defects.

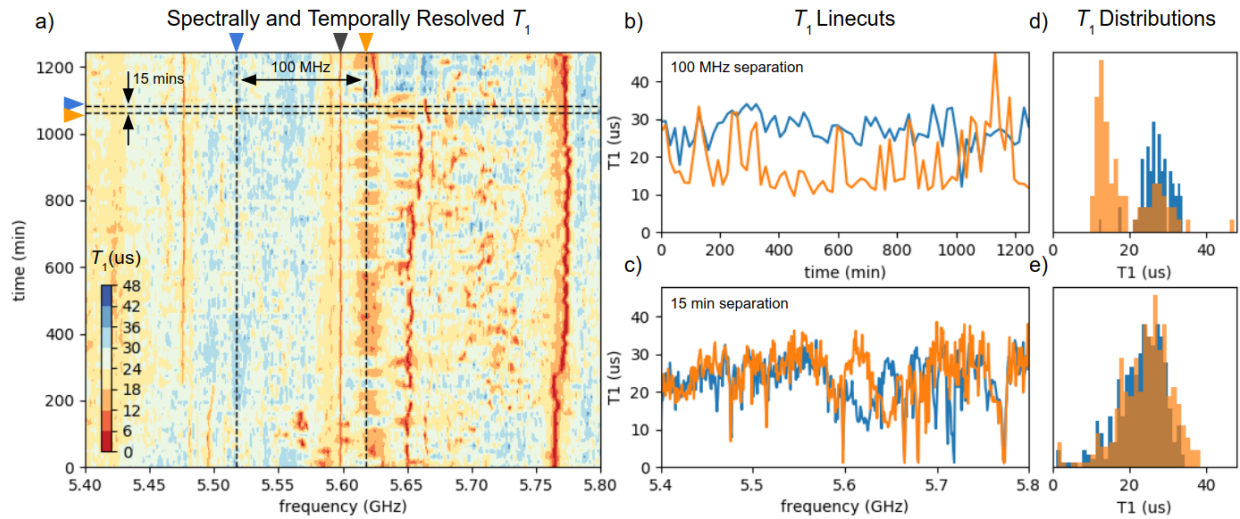

**Figure 2. Spectrally and Temporally Resolved $T_1$.** This dataset comprises 31,278 $T_1$ measurements, spanning 400 MHz and 1200 minutes with 1 MHz and 15 minute step sizes, respectively. Most regions of strong relaxation are consistent with defect-induced relaxation. The resonance near 5.6 GHz, which is indicated with a black arrowhead, is due to bleedthrough of our microwave carrier. Two linecuts at (b) constant frequency and (c) at constant time. The data are taken at the dashed lines in (a), which are color coded to the linecuts with arrowheads. These linecuts show that $T_1$ can fluctuate by an order of magnitude on 15 minute timescales and 5 MHz frequencies. (d) and (e) $T_1$ distributions corresponding to the linecuts in (b) and (c).

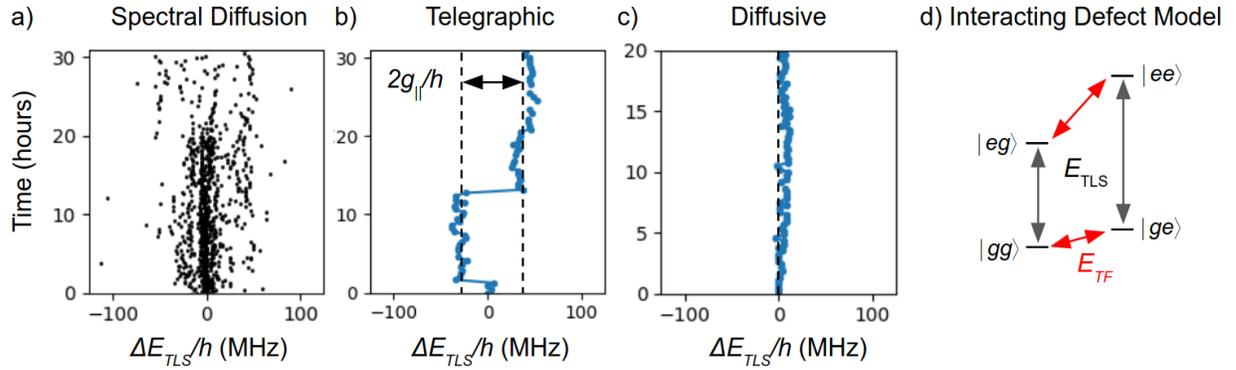

**Figure 3: Defect Spectral Diffusion.** (a) Spectral diffusion of the transition frequencies of 13 defects, with $\Delta E_{TLS}(t) \equiv E_{TLS}(t) - E_{TLS}(0)$. The individual trajectories exhibit both telegraphic and diffusive dynamics (see Fig. S2). (b) Telegraphic spectral diffusion is characterized by abrupt jumps in transition frequency. (c) Diffusive spectral diffusion is characterized by a continuous drift in transition frequency. (d) Energy level diagram of a TLS defect coupled to a thermal-fluctuator (TF) defect. TLS and TF transitions are shown with black and red double arrows, respectively. As the TF thermally transitions between its energy eigenstates, the TLS transition frequency jumps by $2g_{\parallel}/h$, which is observed as telegraphic spectral diffusion. Diffusive dynamics can emerge in the bath limit of this model where a single TLS is coupled to many TFs.

**Supplementary Materials for "Fluctuations of Energy-Relaxation Times in Superconducting Qubits"**

P. V. Klimov[1], J. Kelly[1], Zijun Chen[1], M. Neeley[1], A. Megrant[1], B. Burkett[1], R. Barends[1], K. Arya[1], B. Chiaro[3], Yu Chen[1], A. Dunsworth[3], A. Fowler[1], B. Foxen[3], C. Gidney[1], M. Giustina[1], R. Graff[1], T. Huang[1], E. Jeffrey[1], Erik Lucero[1], J.Y. Mutus[1], O. Naaman[1], C. Neill[1], C. Quintana[1], P. Roushan[1], Daniel Sank[1], A. Vainsencher[1], J. Wenner[3], T.C. White[1], S. Boixo[2], R. Babbush[2], V. N. Smelyanskiy[2], H. Neven[2], John M. Martinis[1]

1. Google, Santa Barbara, California 93117
2. Google, Los Angeles, California 90291
3. University of California, Santa Barbara, California 93117

Sections
- S1. Sample
- S2. Additional $T_1$ Data
- S3. Fitting Relaxation Resonances
- S4. Defect Relaxation Resonances
- S5. Spurious Relaxation Resonances
- S6. Qubit Frequency Calibration
- S7. Defect-Qubit Coupling Strengths
- S8. Defect-Defect Coupling Strengths
- S9. Spectral Diffusion Simulations
- S10. Consolidated Defect Data
- S11. References

S1. Sample
Our sample is a hybrid flip-chip device in which a qubit chip is bump bonded to a control-wiring chip. Our qubits are Xmon transmons with a design similar to that described in refs. [1, 2, 3, 17]. The qubit capacitor and control wiring were patterned by depositing Al in an electron beam evaporator and then etching with a $BCl_3 + Cl_2$ plasma in an ICP etcher. The SQUID Josephson junctions were patterned by shadow-evaporating Al in an electron beam evaporator, with the native $AlO_x$ serving as the junction dielectric. The sample was measured in a dilution refrigerator at a temperature of *T = 15* mK with a microwave control and readout circuit similar to that described in past reports [1].

S2. Additional $T_1$ Data
Additional $T_1$ datasets taken on five qubits on the same chip are presented in Fig. S1. Although we do not present additional data, we note that we have observed similar spectral diffusion dynamics on different chips, with both nominally identical and different design and fabrication parameters. We are thus confident that we are not observing anomalous behavior.

S3. Fitting Relaxation Resonances

We fit spectroscopic energy-relaxation data extracted from Fig. S1 to the energy-relaxation model $\frac{1}{T_1(f)} = \Sigma_i \frac{2(g_i/h)^2 \Gamma_i}{(\Gamma_i/2\pi)^2 + (f_i - f)^2} + \Gamma_{1,Q}$ (Fig. S2a) [17]. This model comprises a sum of Lorentzians, each of which corresponds to resonant energy-relaxation into a coupled quantum system, and a constant background relaxation rate $\Gamma_{1,Q}$. Each Lorentzian is parameterized by the coupled system's resonance frequency $f_i$, it's transverse coupling strength to the qubit $g_i$, and the total decoherence rate $\Gamma_i = \frac{\Gamma_{1,i}}{2} + \Gamma_{\phi,i} + \frac{\Gamma_{1,Q}}{2} + \Gamma_{\phi,Q}$, where $\Gamma_1$ and $\Gamma_\phi$ are the energy-relaxation and dephasing contributions, respectively. In the vicinity of energy-relaxation resonances, the qubit's and coupled systems' dephasing rates are relevant, since they contribute to their respective systems' spectral bandwidths. The qubit contribution to $\Gamma_i$ is $\Gamma_Q = \frac{\Gamma_{1,Q}}{2} + \Gamma_{\phi,Q} \sim 0.70\ MHz$ over our measurement range. This energy-relaxation model is valid in the coupling regime $\Gamma_{1,i} > 2\pi g_i/h > \Gamma_{1,Q}$, which is true for all defect resonances considered here.

## S4. Defect Relaxation Resonances

We identified 13 defects in this study. Their best-fit $\Gamma_i$ and $g_i$ parameters are presented in Fig. S2b. The coupling strengths $g_i$ are discussed further in Section S7. The trajectories of these defects' transition frequencies are consolidated in Fig. S4a. The transition frequency for each defect was estimated as the frequency at the maximum decay rate in a spectral window centered on the defect's trajectory. This method proved to be accurate at scale, returning results similar to the more precise but more manual fitting approach that was used to extract the trajectories presented in main text Figs. 2b and 2c.

## S5. Spurious Relaxation Resonances

Each one of our qubits has a time-independent periodic relaxation resonance every ~180 MHz. These are marked with double arrows in Fig. S1a. Since these resonances are shifted by up to 30 MHz between qubits, it is unlikely that they are box or chip modes. We ascribe them to resonances in our control lines with a simple calculation, which we have confirmed with more precise SPICE simulations (not shown). The length of the coaxial cable between our mix-plate filters and our chip is $L$ = 23" and the dielectric constant of it's teflon dielectric is $\varepsilon_r$ = 2.1. We thus expect a half-wave resonances every $f_r \sim \frac{c}{2L\sqrt{\varepsilon_r}} = 177\ MHz$, which closely matches the periodicity of the relaxation resonances that we observe. We see another spurious relaxation resonance at ~5.6 GHz for all qubits. This is due to bleed-through of the microwave carrier used in qubit control.

## S6. Qubit Frequency Calibration

Before running an experiment on a qubit, we establish a calibration curve that maps from the voltage that we apply to that qubit's SQUID flux-bias line to that qubit's frequency. We do this by measuring the qubit frequency at many flux biases and then fitting the resulting data to the standard transmon energy-flux dispersion [32]. It is important to verify that this calibration stays constant in time since spectral diffusion of the qubit frequency could be misinterpreted as spectral diffusion of TLS resonances. We verify the stability of this calibration curve during *each* spectroscopic $T_1$ measurement through several independent frequency standards, described below.

First, we interleave a qubit frequency measurement with each $T_1$ spectroscopy measurement. This measurement is taken every ~15 minutes at the qubit initialization frequency (e.g. 5.5 GHz for q0_2; data in Fig. 2a). We find that the measured qubit frequency is within ±1 MHz of the desired frequency over ~25 hours of measurement time. This data, consolidated for all measurements taken during Fig. 2a and Fig. S1, is presented in Fig. S3.

Second, we monitor the bleedthrough of our microwave carrier (see Section S5), which can be seen as a resonance in each spectroscopic $T_1$ trace and serves as an excellent frequency standard. This resonance is near 5.6 GHz, and is measured every ~15 minutes. We see that it is stable to within ±1 MHz - limited by our measurement resolution - over ~25 hours of measurement time. Note that we do not have this standard for q0_1 and q1_3, since the carrier is outside of the measured spectrum (see Fig. S1).

Finally, we monitor several control-line modes (see Section S5), which can be seen as periodic resonances in each spectroscopic $T_1$ trace. These resonances occur with ~180 MHz periodicity, and are each measured every ~15 minutes. We expect these resonances to be frequency-stable as they are geometrically defined electromagnetic modes. We find that these modes are also stable to within ±1 MHz - limited by our measurement resolution - over ~25 hours of measurement time.

These measures independently establish that our qubit frequency calibration curve is stable to within ±1 MHz over ~25 hours of measurement time. We are thus confident that qubit frequency drift plays a negligible role in our data and its interpretation.

### S7. Defect-Qubit Coupling Strengths:

The transverse defect-qubit coupling strength $g$ is given by the interaction of the defect's electric dipole with the qubit's electric field $g = \frac{2ed}{x}\sqrt{\frac{hf}{2C_q}}$ [19, 28]. Here $e$ is the electron charge, $d$ is the defect's electric dipole length, $C_q$ is the qubit self-capacitance, $h$ is Planck's constant, $f$ is the qubit and defect frequency, and $x$ is the length of the electric field line. The angular dependence of the defect's dipole with respect to the qubit's electric field has been dropped, so this expression gives the maximum coupling strength.

We now compute order-of-magnitude estimates for reasonable coupling strengths in our architecture. In these estimates, we take *d* = 1 Å, $C_q$ = 75 fF and *f* = 5.5 GHz. Depending on whether we are considering defects within the qubit capacitor, near the Josephson junction, or within it, we use *x* = 20 um, 1 um, 2 nm, respectively. The coupling strengths in these regions range from 10 kHz to 250 kHz to 100 MHz. Since the coupling strengths that we measure range from 100 kHz to 1 MHz (Fig. S2b), we believe that the TLS defects that we observe are located either in the qubit capacitor or near the Josephson junction, but probably outside of the junction.

This calculation is consistent with more sophisticated simulations of qubit-defect coupling strengths in our architecture [17].

S8. Defect-Defect Coupling Strengths:
In the main text, we analyze the TLS-TF interaction Hamiltonian $\hat{H}_{zz} = \frac{1}{2} g_{zz} \hat{\tau}_{z,TLS} \hat{\tau}_{z,TF}$. The coupling strength $g_{zz}$ generally comprises elastic and electrical components, but we ignore the elastic component, which is expected to be small [24]. The coupling strength $g_{zz}$ is therefore $\frac{g_{zz}}{2} = \frac{1}{4\pi\varepsilon_0 \varepsilon_r r^3}(p_{1\perp} p_{2\perp} - 2 p_{1\parallel} p_{2\parallel})$, where $p_1$ and $p_2$ are the TF and TLS dipole moments, $r$ is the distance between them, $\varepsilon_0$ is the vacuum permittivity, $\varepsilon_r$ is the relative permittivity of the host material (assumed 10), and the subscripts $\perp$ and $\parallel$ denote normal and parallel projections of the dipole moments onto the vector joining them.

In the interacting defect model, the magnitude of telegraphic hops is $2g_{\parallel} = 2g_{zz}(\frac{\varepsilon_{TLS}}{E_{TLS}} \frac{\varepsilon_{TF}}{E_{TF}})$. We use this expression to estimate the distance between TLS and TF pairs in our experiments, and to compute the coupling strengths between them in our simulations. When performing these computations, we estimate that $g_{zz} \sim g_{\parallel}$. This is reasonable for TFs, since we typically see telegraphic jumps on timescales of minutes or hours, which implies that $\Delta_{TF}$ is small and that $\varepsilon_{TF} \sim E_{TF}$. We cannot determine how accurate this approximation is for TLS defects, since we cannot measure $\Delta_{TF}$.

S9. Spectral Diffusion Simulations
We consider a 3 nm x 1 μm x 1 μm cuboid, which is representative of interfacial dielectrics in our qubit circuit. We place a TLS at the cuboid center and then randomly populate it with TF defects at a variable density that will be defined later. The TLS and each TF are associated with a dipole moment randomly sampled from the probability distribution $\rho_0 \sqrt{1 - (p/p_{max})^2}$, where $p_{max}$ = 1.5 eÅ [17]. Each TF is assumed to be at a random angle with respect to the TLS. From each TLS-TF pair, we compute a $g_{\parallel}$ coupling strength based on their dipole moments and their relative angles (see Section S7).

Each TF is assumed to have a telegraphic flip rate $\Gamma$ that is sampled from a distribution $\propto \Gamma^{-1}$ [24], which is consistent with our data. For normalization, we constrain the domain of the distribution to ($\Gamma_{min}$, $\Gamma_{max}$). The upper bound is chosen $\Gamma_{max}$= *1/δt*, where *δt* = 15 mins is the temporal resolution of our $T_1$ data. Larger telegraphic flip rates would manifest as parallel TLS lines in our data, which we have not seen across many samples. The lower bound is chosen to be $\Gamma_{min}$= *1 / 2t*$_{sim}$ where $t_{sim}$ = 30 hours is the duration of our simulation and measurements. At and below this flip rate, we do not expect to see any telegraphic hops in our data. We assume that the flip rate is equivalent for excitation and relaxation, which is approximately true for most TFs by definition and is sufficient to get a qualitative understanding of the dynamics.

We simulate spectral diffusion dynamics over $t_{sim}$=30 hours, taking *δt* = 15 minute time steps, to match our measurements. At *t* = 0, we initialize each TF into a random state and initialize the relative TLS transition frequency to $\Delta E_{TLS}/h$ = 0 MHz. Then, for each of $t_{sim}/\delta t$ steps, we allow each TF to flip with its respective probability $\Gamma \delta t$. Each flip event shifts the TLS transition frequency by $2g_\parallel$, leading to spectral diffusion. Spectral diffusion trajectories simulated in this way, at a variable TF density, are presented in Fig. S4b.

To compare simulation with experiment, we simulate the diffusivity of samples of 13 simulated trajectories at densities in the range ~$10^2$ - $10^5$ GHz$^{-1}$ µm$^{-3}$ (Fig. S5). We do not simulate above a density of ~5.0 x$10^4$ GHz$^{-1}$ µm$^{-3}$, at which point the average TF-TF spacing per GHz is ~25 nm, and so TF-TF interactions may start to play an important role in their dynamics. From the simulations, we see that up to a density of $10^3$ GHz$^{-1}$ µm$^{-3}$, most of the dynamics are too small to resolve in our measurements (< 1 MHz). At a density of ~$10^4$ GHz$^{-1}$ µm$^{-3}$, our experimentally measured diffusivity $D$ = 2.5 $\mp$ 0.1 MHz (hour)$^{-\frac{1}{2}}$ is consistently achieved, and the simulated spectral diffusion trajectories qualitatively match our data (Fig. S4). This is ~10x higher than the 0.2 - 1.2×$10^3$ GHz$^{-1}$ µm$^{-3}$ (0.5 − 3 × $10^{20}$ eV$^{-1}$ cm$^{-3}$) densities that are typically quoted for bulk dielectrics [24].

S10. Consolidated Defect Data:
All defect parameters are consolidated in Table S1. Here we summarize how they were computed. We extract $g_i$, and $\Gamma_i$ from Lorentzian fits to spectroscopic relaxation data as described in Section S2. We estimate $g_\parallel/h$ from trajectories where clear telegraphic hopping is observed as half of the jump amplitude as described in Section S8. We estimate $(\Gamma_{e \to g} + \Gamma_{g \to e}) / 2$ as the total number of jumps per measurement time, which implicitly assumes that $\Gamma_{e \to g} \sim \Gamma_{g \to e}$ (i.e. $E_{TF} < k_B T$). This is a very rough estimate for trajectories where few telegraphic jumps are observed. We compute $E_{TF}/k_B T = \ln(\Gamma_{e \to g}/\Gamma_{g \to e})$ for trajectories where many telegraphic jumps are observed.

| qubit | defect | $g_i/h$ (MHz) | $\Gamma_i$ (MHz) | $g_\parallel/h$ (MHz) | $(\Gamma_{e\to g} + \Gamma_{g\to e})/2$ (hr$^{-1}$) | $E_{TF}/k_B T$ |
|---|---|---|---|---|---|---|
| q1_3 | d0 | 0.131 ∓ 0.012 | 2.38 ∓ 8.18 | 1 | 2.30 | 0.18 |
|  | d1 | 0.163 ∓ 0.012 | 0.69 ∓ 1.62 | - | - | - |
| q1_2 | d0 | 0.594 ∓ 0.010 | 3.77 ∓ 1.39 | 12 | 0.05 | - |
|  | d1 | 0.179 ∓ 0.168 | 64.47 ∓ 13.85 | 12 | 1.10 | 0.99 |
|  | d2 | 0.258 ∓ 0.112 | 0.96 ∓ 1.01 | 16 | 0.05 | - |
|  | d3 | 0.187 ∓ 0.004 | 5.55 ∓ 0.26 | - | - | - |
|  | d4 | 0.237 ∓ 0.003 | 2.66 ∓ 0.27 | 4 | 0.10 | - |
| q0_1 | d0 | 0.426 ∓ 0.010 | 23.84 ∓ 1.42 | - | - | - |
| q0_3 | d0 | 0.248 ∓ 0.006 | 11.19 ∓ 0.69 | 15, 30 | 0.07, 0.03 | - |
| q0_4 | d0 | 0.401 ∓ 0.033 | 8.26 ∓ 1.51 | 20, 3 | 0.03, 0.05 | 0.28 |
|  | d1 | 0.442 ∓ 0.049 | 11.60 ∓ 0.33 | - | - | - |
|  | d2 | 0.297 ∓ 0.045 | 4.87 ∓ 2.56 | - | - | - |
|  | d3 | 0.223 ∓ 0.084 | 10.39 ∓ 9.32 | - | - | - |

**Table S1**. **Defect Parameters.** The qubit and defect columns correspond to the labels in Fig. S1. See Section S10 for a summary of how these parameters were calculated.

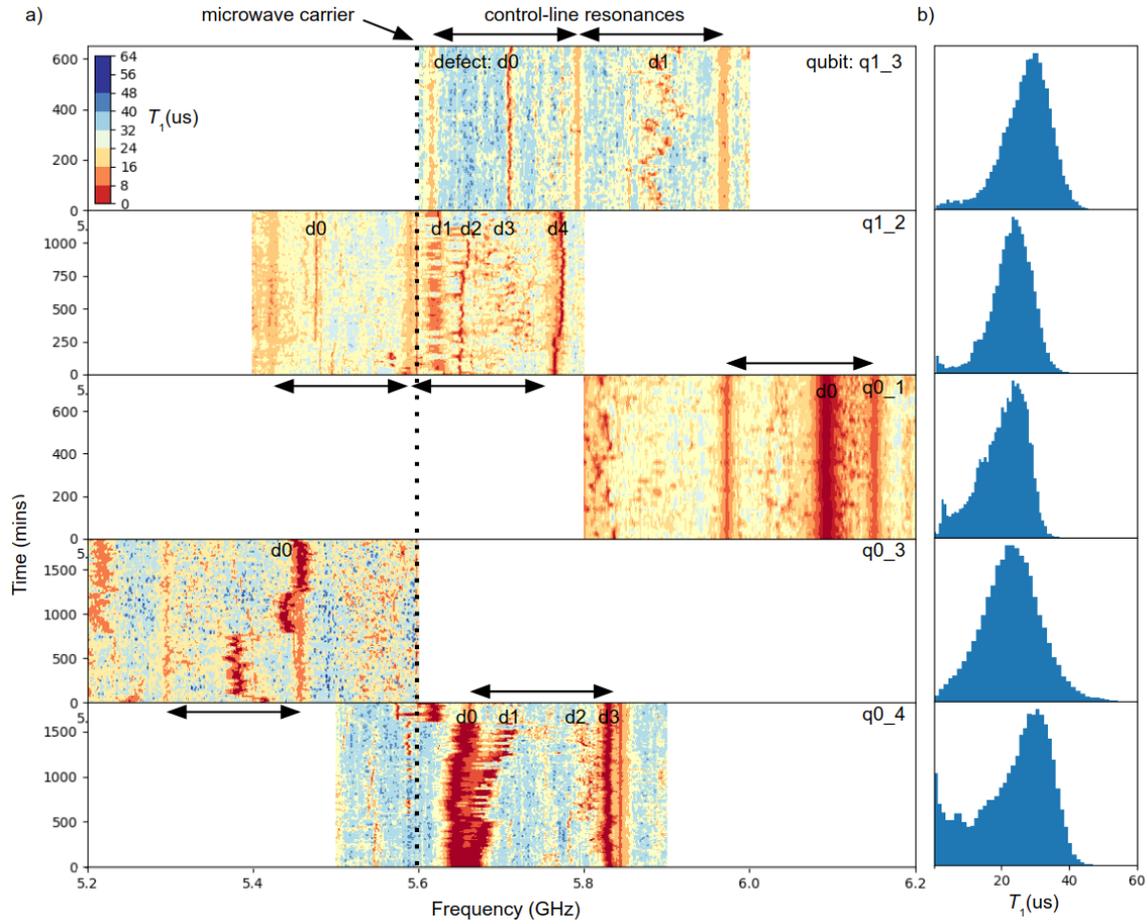

**Figure S1. Additional Data.** (a) Spectrally and temporally resolved $T_1$ taken on 5 qubits on the same chip. We attribute many relaxation resonances to defects, and plot the individual defect trajectories in Fig. S3a. Each panel has labels indicating the qubit (e.g. q1_3) and analyzed defects (e.g. d0). Parameters extracted for these defects are presented in Table S1. We attribute periodic relaxation resonances to resonances in our control lines (double arrows) and the sharp resonance near 5.6 GHz to bleedthrough of our microwave carrier (dashed line). (b) $T_1$ distributions corresponding to the datasets presented in (a).

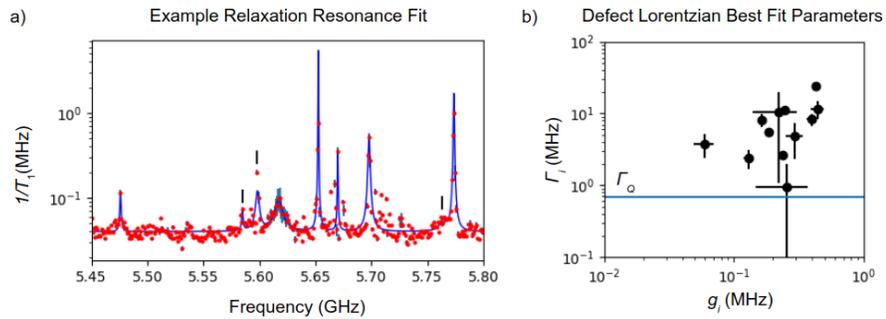

**Figure S2. Fitting Relaxation Data.** (a) Example fit of our energy-relaxation model to a $1/T_1$ energy-relaxation scan. Black vertical bars mark spurious non-defect-related resonances. (b) The best fit parameters and corresponding 68% confidence intervals for the 13 defects extracted from the data in Fig. S1. $\Gamma_Q$ is the contribution to $\Gamma_i$ from qubit energy-relaxation and dephasing.

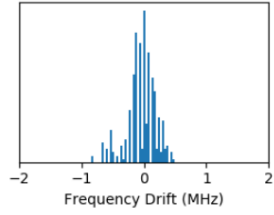

**Figure S3: Qubit Frequency Stability.** Distribution of qubit-frequency drifts calculated from qubit frequency measurements taken interleaved with each spectroscopic $T_1$ measurement presented in Fig. 2a and Fig. S1a. Frequency drift is defined as the difference between the measured qubit frequency and the initially calibrated qubit frequency. This data indicates that our qubit-frequency calibrations are stable to within ±1 MHz over the course of our measurements.

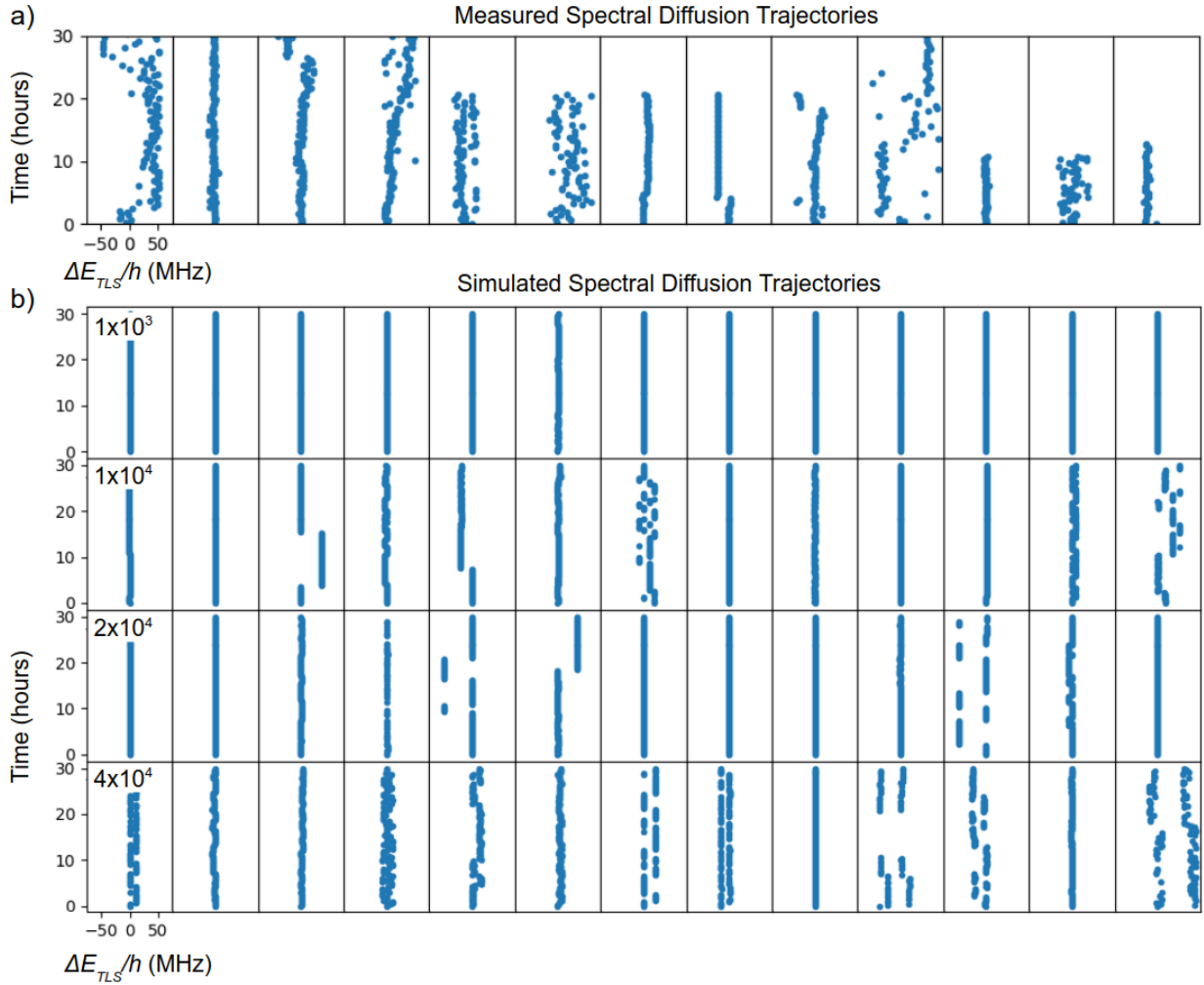

**Figure S4. Measured and Simulated Spectral Diffusion Trajectories.** (a) Spectral diffusion trajectories of the 13 defects extracted from the data presented in Fig. S1. See Section S4 for a description of how these trajectories were extracted. (b) Simulated spectral diffusion trajectories at densities ranging from 1 to 4 x $10^4$ GHz$^{-1}$ µm$^{-3}$. All plots share the same axes. The dynamics that we observe are qualitatively reproduced near $10^4$ GHz$^{-1}$ µm$^{-3}$.

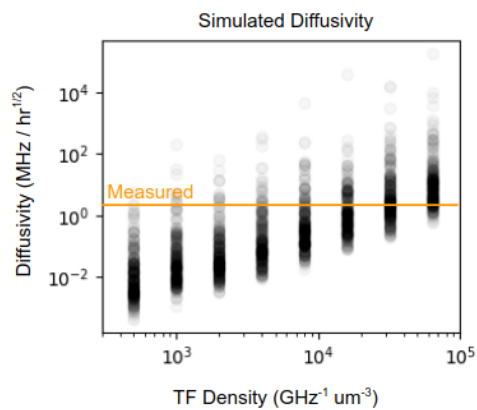

**Figure S5. Simulated Diffusivity versus TF Density.** Each point represents the diffusivity of 13 simulated spectral diffusion trajectories. There are 300 points per density. Our experimentally measured diffusivity of $D = 2.5 \mp 0.1$ MHz (hour)$^{-½}$ is consistently achieved at TF densities near $10^4$ GHz$^{-1}$ μm$^{-3}$.